\documentclass[reprint,aip,apl,floatfix,letterpaper]{revtex4-1}
\usepackage{graphicx}
\usepackage[product-units = power, list-units = single]{siunitx}
\usepackage[utf8]{inputenc}
\usepackage[T1]{fontenc}
\usepackage{natbib}
\usepackage{footnote}
\usepackage{epstopdf}

\begin{document}

\title{Exchange coupling of a perpendicular ferromagnet to a half-metallic compensated ferrimagnet via a thin hafnium interlayer}
\author{Kiril Borisov}
\author{Gwena\"el Atcheson}
\author{Gavin D'Arcy}
\author{Yong-Chang Lau}
\altaffiliation[Present Address: ]{Department of Physics, The University of Tokyo, Bunkyo, Tokyo 113-0033, Japan}
\author{J.\,M.\,D.\,Coey}
\author{Karsten Rode}
\affiliation{CRANN, AMBER and School of Physics, Trinity College,
Dublin 2, Ireland}
\date{\today}

\begin{abstract}
  A thin Hafnium film is shown to act both as an effective diffusion barrier for manganese at a
  thickness of \SI{0.7}{\nano\metre}, and as an effective exchange
  coupling layer in a sandwich structure with perpendicular magnetic
  anisotropy. The magnetic layers are Co$_{20}$Fe$_{60}$B$_{20}$ and the low
  moment ferrimagnet Mn$_2$Ru$_x$Ga (MRG).
  The coupling changes sign at the compensation temperature of MRG and the exchange energy
  reaches \SI{0.11}{\milli\joule\per\metre\squared} for the thinnest Hf
  interlayers. Ruthenium, the usual metal of choice for coupling ferromagnetic layers in
  thin film heterostructures, cannot be used with the zero-moment half metal
  MRG because of Ru interdiffusion.
  Due to its large coercivity near compensation, the MRG can act as
  an effective source of exchange pinning.
\end{abstract}

\maketitle


Spin electronics has influenced the information revolution
through technologies such as affordable high-density magnetic storage and
high-speed nonvolatile magnetic memory, taking advantage of the giant
magnetoresistance or tunneling magnetoresistance (GMR or TMR) effects to read
and store information. In the magnetic thin film stacks employed as field
sensors or magnetic switches, it is useful to be able to control the magnetic
coupling between adjacent layers. Direct contact between an antiferromagnetic
and a ferromagnetic layer can lead to exchange bias of the latter, enabling it
to serve as a pinned or reference layer in any spin valve sandwich structure.
Stacks may also make use of the interlayer exchange coupling \cite{Parkin1990,Parkin1991} to form synthetic antiferromagnets (SAFs) where two
ferromagnetic layers are coupled antiferromagnetically via a thin layer of
nonmagnetic metal in a structure that creates no stray field. The exchange
coupling can be further engineered to form structures that enable demonstration
of ultrafast chiral domain wall motion \cite{Yang2015} and deterministic
spin orbit torque-\-induced switching of a perpendicular ferromagnet without an external field \cite{Lau2016}. The sign of the
exchange usually oscillates with the nonmagnetic spacer thickness, and Ru has
been found to be the most effective \cite{Parkin1990}. A peak in the antiferromagnetic
coupling is found for a ruthenium thickness of \SI{0.9}{\nano\metre}, where the
exchange energy reaches \SI{0.1}{\milli\joule\per\metre\squared}. An early
systematic study by Parkin found no coupling between 3$d$ ferromagnetic layers
for the $d^2$ transition metals Ti, Zr or Hf \cite{Parkin1991}.

An alternative to using an exchange-biased SAF to pin the reference layer is to
exchange-couple it to a low-moment uniaxial ferrimagnet that is close to
compensation. This could produce the necessary exchange bias, whilst adding
magnetic mass to stabilise ultra-thin CoFeB layers in perpendicular magnetic
tunnel junctions (MTJs) \cite{Ikeda2010}. The uniaxial anisotropy $K_1$ ensures that, as the
magnetisation $M_s$ of the ferrimagnet approaches zero at the compensation
temperature ($T_{\text{comp}}$), the anisotropy field $2 K_1 / \mu_0M_s$
diverges and the coercivity increases significantly \cite{Nivetha2015}. Thin films of the
recently-discovered compensated ferrimagnetic half-metal Mn$_2$Ru$_x$Ga (MRG)
\cite{Kurt2014} have been shown to exhibit high spin-polarisation and a large coercivity
near compensation \cite{Borisov2016, Nivetha2015}, due to its perpendicular magnetic anisotropy (PMA)
caused by biaxial substrate-induced strain of the cubic Heusler structure. MRG
has a Heusler-type structure, with Mn on $4a$ and $4c$ sites, Ga on
$4b$ sites and Ru on most of the $4d$ sites.\cite{Kurt2014, Betto2015} Some Mn-Ga antisite
substitution is present in the structure \cite{Betto2015}, which leads to a spin gap at the
Fermi level \cite{Zic2016}. Here we show that MRG can be coupled antiferromagnetically
to an adjacent perpendicular CoFeB layer via a thin Hf interlayer, and
the effect can be used to produce exchange pinning.


\begin{figure}
  \centering
  \input{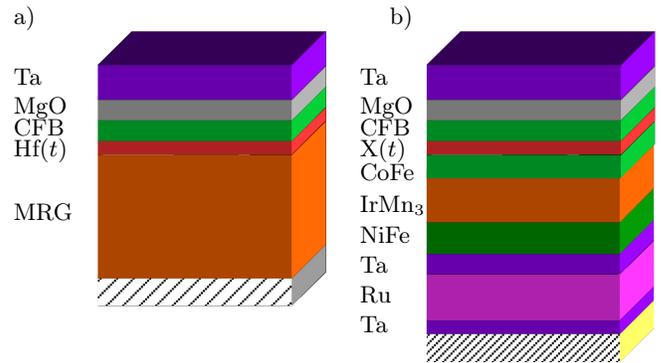}
  \caption{Fully PMA thin film stack used to investigate exchange coupling via Hf in a
    perpendicular ferromagnetic system (a) and an in-plane magnetised reference stack without MRG
    (b), X = Hf or Ru. The Hf thickness, $t$, was varied from \SIrange{0.4}{1.6}{\nano\metre}.}
  \label{fig:stacks}
\end{figure}
The Mn$_2$Ru$_{0.7}$Ga used in this study has $T_{\text{comp}} = \SI{240}{\kelvin}$ so data can be conveniently
collected at temperatures above and below compensation.  Stacks of
MgO//\-MRG(35)/\-Hf($t$)/\-CFB(1)/\-MgO(1.4)/\-Ta(1) [thicknesses in \si{\nano\metre}], where $t = $ \SIlist{0.3;0.4;0.5;0.6;0.7;0.8;0.9;1.1;1.2;1.6}{\nano\metre}, were grown on \SI{10 x 10}{\milli\metre} single-crystal MgO (100)
substrates in Shamrock magnetron sputter deposition tool. Here CFB denotes
Co$_{20}$Fe$_{60}$B$_{20}$ (nominal target composition). 
The MRG was DC co-sputtered from stoichiometric
Mn$_2$Ga and Ru targets, at a substrate temperature of \SI{380}{\degreeCelsius}.\cite{Kurt2014, Nivetha2015} After cooling to
room temperature, Hf and CFB were DC sputtered and, finally, MgO and Ta were RF sputtered. 
Stacks were post-annealed under a vacuum of
$\sim\SI{e-6}{\milli\bar}$ in an applied out-of-plane
magnetic field of \SI{800}{\milli\tesla} at \SI{300}{\degreeCelsius} for \SI{30}{minutes} to enhance the PMA of CFB.
In-plane magnetised reference stacks \cite{Yuasa2007} of composition
\-Ta(5)/\-Ru(10)/\-Ta(5)/\-Ni$_{81}$Fe$_{19}$(5)/\-Ir$_{22}$Mn$_{78}$(10)/\-Co$_{90}$Fe$_{10}$(2.4)/\-X($t$)/\-CFB(3)/\-MgO(1.4)/\-Ta(5), where X~=~Hf or Ru and $t$ = 0.5, 0.7, 0.9 and \SI{1.0}{\nano\metre}, were
deposited on thermally oxidised \SI{20 x 20}{\milli\metre} Si/SiO$_2$ coupons at room temperature.
Reference stacks were post-annealed in an in-plane applied field at \SI{350}{\degreeCelsius} for \SI{60}{minutes} to set the exchange bias at IrMn/CoFe interface.
The stacks are illustrated in \figurename~\ref{fig:stacks} a) and b).

Anomalous Hall effect (AHE) was measured using the 4-point van der Pauw method, with
indium contacts, an applied current of \SI{1}{\milli\ampere} and an applied
magnetic field of up to \SI{1}{\tesla}. Additional Hall measurements up to
\SI{14}{\tesla} were
carried out in a Quantum Design PPMS system. Room temperature magnetisation measurements with magnetic field applied in-plane or perpendicular to the films were carried out using a Quantum Design SQUID with a
maximum field of \SI{5}{\tesla}.

If we are to use MRG, and more generally the Mn-containing Heusler alloys, in an exchange coupling system, we have to consider three
factors when deciding what metal interlayer to use it with: 
Does it prevent interdiffusion between the MRG and the adjacent layers?\cite{Hayakawa2006} 
Does it maintain perpendicular anisotropy of the overlayer?\cite{Liu_AIPadvances2012,Pai2014,Oh2014} Does it
provide sufficient coupling between the MRG and an adjacent ferromagnetic layer?\cite{Radu2012} 
An ideal material will do all three, but the selection is limited. Initially, a number of
candidates were chosen, namely W, Zr, Ru, Mo, Hf, Ta and TiN. Some of
them were effective at blocking diffusion between layers \footnote{The
diffusion profile has been recorded by Time-of-Flight Secondary Ion Mass
Spectroscopy, in preparation.}, which is necessary
to preserve spin-polarisation and device integrity \citep{Hayakawa2006}. Only Mo, Hf and Ta also
maintained the PMA of the CFB overlayer, and of these Hf was the only one to demonstrate strong
effective exchange coupling between MRG and CFB. An order of magnitude smaller effect was found with the
thinnest Ta (\SI{0.4}{\nano\metre}). Ruthenium, the usual material of choice for exchange coupling
ferromagnetic films, is unsuitable due to the lack of PMA in Ru/CFB/MgO heterostructures \cite{Worledge2011} and the incorporation of Ru
into the MRG layer, which alters the interfacial magnetic properties.


The Mn$_2$Ru$_{0.7}$Ga composition used here has a
room-temperature
magnetisation of \SI{40}{\kilo\ampere\per\metre} and coercivity $\mu_0H_c =
\SI{0.7}{\tesla}$. For $t = \SI{0.3}{\nano\metre}$, we found neither
PMA nor exchange coupling of CFB for any of the seven interlayer materials.

\begin{figure}
  \centering
  \input{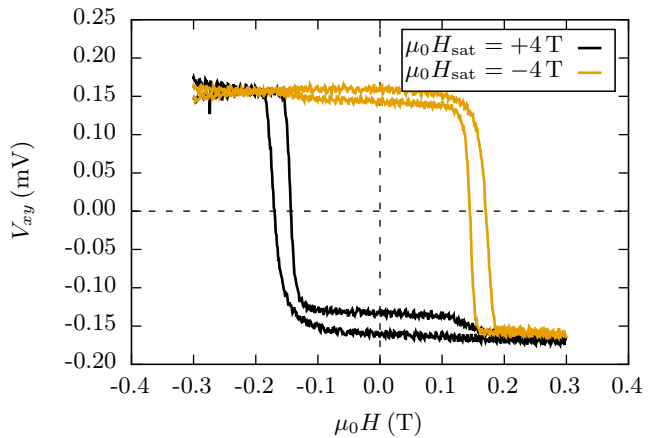}
  \caption{Minor AHE loops for the stack of
    \figurename~\ref{fig:stacks}a) with a \SI{0.4}{\nano\metre} Hf interlayer at
    \SI{300}{\kelvin} after saturation at $\pm\SI{4}{\tesla}$, showing strong effective ferromagnetic exchange coupling of the MRG and CFB layers. Field is applied perpendicular to plane. }
  \label{fig:ahe-gmw}
\end{figure}
\begin{figure}
  \centering
  \input{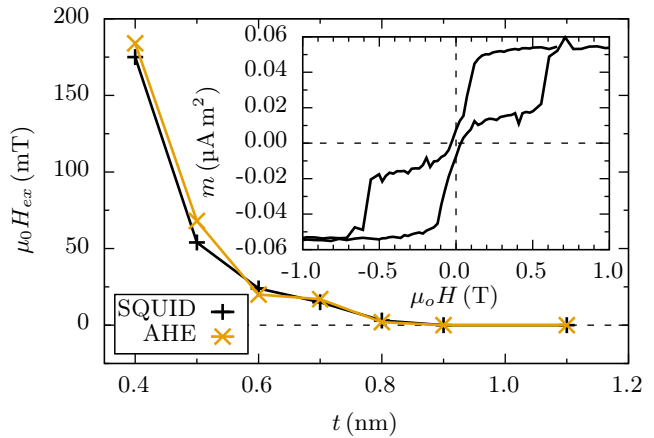}
  \caption{Exchange shifts measured by SQUID or AHE as a function of the
    thickness of the Hf spacer layer. The insert shows SQUID magnetisation loops
    for a stacks of \figurename~\ref{fig:stacks}a) with $t = \SI{0.4}{\nano\metre}$.
    Note that the SQUID loop was measured below $T_{\text{comp}}$ of MRG.}
  \label{fig:squid}
\end{figure}

The effect of the exchange coupling is nicely illustrated for
a Hf interlayer with $t = \SI{0.4}{\nano\metre}$ in \figurename~\ref{fig:ahe-gmw},
where the CFB layer is perpendicular and minor hysteresis loops can be readily
measured using the AHE. The narrow hysteresis loop of the CFB with a coercivity
of \SI{16}{\milli\tesla} is exchange shifted by $\mu_0H_{ex}$~=~\SI{175}{\milli\tesla} in the direction
opposite to the saturation field. These measurements have been repeated
for greater Hf thickness, and $\mu_0H_{ex}$ decreases
monotonically with $t$, falling to zero at $t \approx \SI{1.0}{\nano\metre}$
(\figurename~\ref{fig:squid}).
Note that the exchange shift of the loop in \figurename~\ref{fig:ahe-gmw} is towards
negative fields for ferromagnetic effective interlayer coupling. This is the same as the normal exchange bias effect caused by direct
contact between antiferromagnetic and ferromagnetic layers.

Complete hysteresis loop for $t=\SI{0.4}{\nano\metre}$ is shown in the inset
\figurename~\ref{fig:squid}, and the values of $\mu_0H_{ex}$ measured by AHE and
SQUID magnetometry are compared in \figurename~\ref{fig:squid}. In agreement with the AHE data,
the greatest exchange shift $\mu_0H_{ex}=\SI{174}{\milli\tesla}$ is found for
the thinnest interlayer, with $t=\SI{0.4}{\nano\metre}$, and the shift has
disappeared at $t = \SI{0.9}{\nano\metre}$. The exchange energy $J_{ex}$ is
estimated as $\mu_0H_{ex}M_{\text{CFB}}t_{\text{CFB}}$.\citep{Meiklejohn1956, Meiklejohn1962} Taking $M_{\text{CFB}} =
\SI{0.6}{\mega\ampere\per\metre}$, the maximum value of $J_{ex}$ is
\SI{0.11}{\milli\joule\per\metre\squared}. This is comparable to the
antiferromagnetic exchange coupling for CoFe/CFB layers separated by a Ru interlayer of thickness $t_{\text{Ru}}$~$\approx$~\SI{0.9}{\nano\metre}\cite{Jung2012}. It has to be noted that this thickness corresponds to the second AFM peak of Ru which is easily attainable, while reports on the first AFM peak showing stronger coupling (\SI{2.2}{\milli\joule\per\metre\squared} - \SI{5}{\milli\joule\per\metre\squared}) with $t_{\text{Ru}}$ = \SI{0.4}{\nano\metre}-\SI{0.5}{\nano\metre} are scarce\cite{Parkin1990, Lau2013, Yakushiji2015}.

\begin{figure}
  \centering
  \input{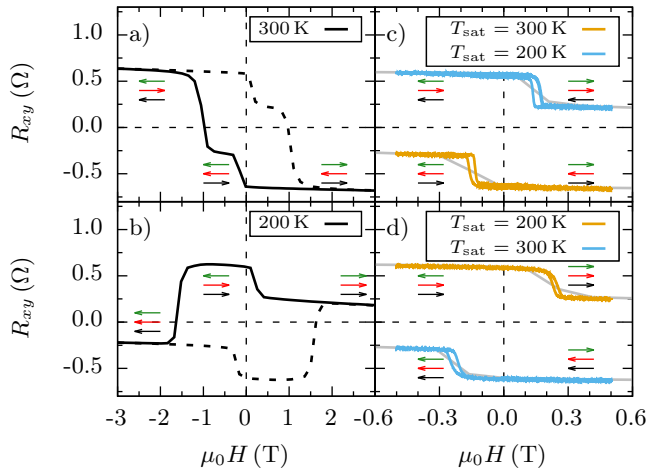}
  \caption{Major AHE loops (left column) for the structure of
    \figurename~\ref{fig:stacks}a) with $t = \SI{0.4}{\nano\metre}$. The field
    sweep $+\rightarrow -$ ($-\rightarrow +$) is drawn in solid (broken) lines. 
    The minor loops measurement protocol (right column) is described in the
    text. Measurements are taken at $T=$ \SIlist{300;200}{\kelvin} ($>$ and
    $<$ $T_{\text{comp}}$) in the top
    and bottom row, respectively. The small arrows indicate the relative
    directions of the CFB (in green), MRG $4c$
    sublattice (red), and net MRG magnetisations (black) during a field sweep
    from positive to negative, with respect to the applied field direction,
    which is \emph{perpendicular} to the sample surface.  The data
    establish that the effective exchange coupling between MRG and CFB changes
    when MRG is saturated below or above $T_{\text{comp}}$.}
  \label{fig:ppms}
\end{figure}
In \figurename~\ref{fig:ppms} a) and b) full AHE loops are recorded above
($\SI{300}{\kelvin}$) and below ($\SI{200}{\kelvin}$) $T_{\text{comp}}$.
The applied field sequence was $+\rightarrow -\rightarrow +$. We
analyse the behaviour in terms of the direction of the CFB magnetisation and the MRG
$4c$ sublattice one, which is the dominant contribution to the MRG electronic
structure at the Fermi level \cite{Betto2015}. At \SI{300}{\kelvin} ($T > T_{\text{comp}}$),
 the $4c$ sublattice is the minority sublattice and hence is antiparallel to the net magnetisation. CFB
switches after passing zero applied field (\figurename~\ref{fig:ppms}a), the coupling is
ferromagnetic with the net MRG moment and antiferromagnetic with the $4c$
moment. At $\SI{200}{\kelvin}$ (\figurename~\ref{fig:ppms}b), the $4c$
sublattice is the majority sublattice, parallel to the net MRG magnetisation.
CFB now switches \emph{before} reaching zero applied field, corresponding to antiferromagnetic
coupling with the net moment \emph{and} with the $4c$ sublattice. The sign of
the effective coupling relative to the net magnetisation changes when crossing
$T_\text{comp}$. Also note that the MRG contribution to the AHE changes
sign when crossing $T_{\text{comp}}$ in accordance with our previous results.\cite{Nivetha2015}

In order to highlight the dependence of the exchange coupling on the sublattice
moment, as
opposed to the net moment, we measured minor loops in the range
$\pm\SI{0.5}{\tesla}$, both above and below $T_{\text{comp}}$
(\figurename~\ref{fig:ppms} c and d) after saturating in a field of
\SI[retain-explicit-plus]{+14}{\tesla}. Two measurements were made in each
case. First after saturation at the measurement temperature, and subsequently
after saturation on the other side of $T_{\text{comp}}$ and heating
(\figurename~\ref{fig:ppms}c) or cooling (\figurename~\ref{fig:ppms}d) back to the
measurement temperature in zero field.
When crossing $T_{\text{comp}}$ in zero field the net MRG magnetization, $M$, changes from positive to negative
while the direction of the $4c$ magnetization, $M_{4c}$, is preserved and
therefore all four possible combinations between $M$ and $M_{4c}$ are accessible.
We find
that the coupling is always antiferromagnetic with respect to $4c$, but the
alignemnt is
ferromagnetic with respect to the net moment above $T_{\text{comp}}$ and
antiferromagnetic below.

\begin{figure}
  \centering
  \input{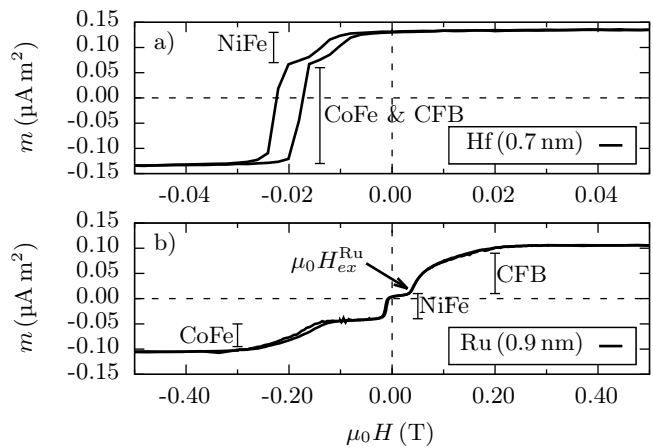}
  \caption{SQUID magnetometry of the in-plane reference stack of \figurename~\ref{fig:stacks} b). Panel (a): Strong ferromagnetic coupling with $t_{\text{Hf}} = \SI{0.7}{\nano\metre}$. The exchange bias is $\mu_{0}H_{ex}^{\text{IrMn}} \approx \SI{22}{\milli\tesla}$. Panel (b): Antiferromagnetic coupling with $t_{\text{Ru}} = \SI{0.9}{\nano\metre}$. The exchange shift is $\mu_{0}H_{ex}^{\text{Ru}} \approx \SI{33}{\milli\tesla}$ which is indicated with an arrow. The latter is a pessimistic estimate.}
  \label{fig:reference}
\end{figure}
Finally, we show in \figurename~\ref{fig:reference} the SQUID loops for the
in-plane reference stack of \figurename~\ref{fig:stacks}b) where a CoFe layer is
separated by a hafnium or ruthenium interlayer from the upper CFB layer. The data
show clear evidence of antiferromagnetic exchange coupling for Ru(\SI{0.9}{\nano\metre}),
and strong ferromagnetic coupling when the spacer is Hf(\SI{0.7}{\nano\metre}), in contrast with
early reports\cite{Parkin1991}. Comparing the exchange coupling energy obtained with
IrMn and Ru in our reference stacks, the exchange shifts are $\mu_{0}H_{ex}^{\text{IrMn}}~\approx~\SI{22}{\milli\tesla}$ and $\mu_{0}H_{ex}^{\text{Ru}}~\approx~\SI{33}{\milli\tesla}$, hence the exchange coupling energies are
$J_{ex}^{\text{IrMn}}~\approx~\SI{0.04}{\milli\joule\per\metre\squared}$ and
$J_{ex}^{\text{Ru}}~\approx~\SI{0.15}{\milli\joule\per\metre\squared}$.


The strength of the coupling via a \SI{0.7}{\nano\metre} Hf interlayer is only
about \SI{20}{\percent} of that for a similar thickness of Ru, but for a \SI{0.4}{\nano\metre}
interlayer it is the same as that of the Ru ($t_{\text{Ru}}$ = \SI{0.9}{\nano\metre}),
and it shows no oscillations of sign, neither in the MRG-based stacks nor in the reference
stacks. This suggests a coupling mechanism that is different to the accepted
one for Ru interlayers.\cite{Bruno1995}

Unlike ruthenium\cite{Parkin1990}, where the sign of the exchange coupling oscillates
with interlayer thickness with a period of about \SI{1.1 }{\nano\metre},
reminiscent of the RKKY interaction, the exchange coupling through hafnium is
always antiferromagnetic in the case of MRG/\-Hf/\-CFB and ferromagnetic for the
symmetric CoFe/\-Hf/\-CFB stack, and it decreases monotonically to zero with
increasing interlayer thickness. The coupling is strongest for the very thin
\SI{0.4}{\nano\metre} films that are only two atomic layers thick. A possible
mechanism for the coupling is hybridization of the
unpolarized Hf $6s^25d^2$ conduction electrons with the spin polarized $4c$
sub-band at the Fermi surface of MRG. This polarizes the first few layers of
Hf, rather as the $5d6s$ bands of the rare-earth metals are polarized by the
$4f$ core. Coupling between an atom with a nearly full $d$ band (Co) and one
with a nearly empty $d$ band (Hf) is antiferromagnetic.\cite{Campbell1969, Campbell1972} The effective
exchange coupling is ferromagnetic when $4c$ is the minority sublattice above
$T_{\text{comp}}$ and antiferromagnetic when $4c$ becomes the majority
sublattice below $T_{\text{comp}}$. We saw from \figurename~\ref{fig:ppms} that the
exchange coupling changes sign at $T_{\text{comp}}$  in just the sense that is
predicted. For the symmetric reference stack CoFe/\-Hf/\-CFB these two
antiferromagnetic interactions leads to an overall ferromagnetic exchange as
observed. Similar effect has been previously observed in
$L$1$_0$-MnGa/Fe$_{1-x}$Co$_{x}$\cite{Ma2014} as a function of $x$
which was attributed to the reversal of the interfacial spin polarisation.

Another possible mechanism is exchange coupling via
pinholes in the thin interlayer which could produce the same effect, with a sign
change at compensation. The \SI{0.4}{\nano\metre} Hf layer is unlikely to be
entirely defect-free but this explanation is implausible for several reasons.
First it should apply to any interlayer with pinholes. Hf is the only one of
the seven metals studied to exhibit the strong exchange
coupling. Furthermore, no exchange was observed for
$t_{\text{Hf}}=\SI{0.3}{\nano\metre}$ where the case for pinholes should
have been even stronger. Finally when CFB is deposited directly on MRG, no
exchange is observed. We can therefore discount this explanation.


Exchange bias has previously been demonstrated by direct coupling of a
compensated ferrimagnet with a ferromagnet\cite{Ungureanu2010, Romer2012}, but
the advantages of our approach are the MRG high ordering temperature
($T_{\mathrm{C}}$ $\approx$ \SI{550}{\kelvin})\cite{Kurt2014} and the broad compensation temperature tunability (\SI{2}{\kelvin}-\SI{400}{\kelvin})\cite{Nivetha2015}. The use of a single layer of MRG or
some other near-zero-moment ferrimagnet close
to compensation can in principle replace the standard bottom SAF reference
stack for MTJ with PMA, e.g. buffer layer/\-[Co/Pt]$_n$/\-Ru/\-[Co/Pt]$_n$/\-Ta
or W/\-CFB,\cite{Cuchet2013,Yakushiji2015} provided the uniaxial anisotropy is
finite so that the anisotropy field diverges when $M_s \rightarrow 0$ as
discussed above.
The
difficulty with using MRG, or any Mn-based ferrimagnet in direct contact with a ferromagnet or a
tunnel barrier such as MgO is manganese diffusion into the overlayer.\cite{Borisov2016, Kubota2011, Hayakawa2006}
Here we have demonstrated the feasibility of a simple bcc-textured MRG/\-Hf/\-CFB reference structure
that can be grown directly on MgO as an alternative. A fully bcc-textured MTJ stack with robust reference layer can therefore be realised. Furthermore, the use of MRG with high spin polarisation in conjunction with a Hf interlayer with relatively high spin transparency \cite{Pai2014} may also be advantageous for enhancing the spin-torque efficiency of the MTJ.


The research on Mn-based Heusler ferrimagnets have instigated considerable scientific interest recently due 
to broad range of magnetic properties\cite{Kurt2014, Nayak2015, Betto2017},  generation of sub-THz
oscillation\cite{Awari2016, Mizukami2016} and successful integration in magnetic tunnel
junction\cite{Borisov2016, Jeong2016}.
We have previously demonstrated that the compensated half-metallic Heusler
alloy MRG combines the advantageous properties of both ferro- and
anti-ferromagnets without their limitations. It can be used as a source of
spin-polarised currents, is immune to external fields and produces no stray
field of its own. In this work we have demonstrated yet another use: as a
reference layer in a spin electronic stack, either alone or as a pinning
material, thereby greatly reducing the number of individual layers in a stack.

  K.B.  acknowledges financial support from Science
  Foundation Ireland within SSPP (11/SIRG/I2130). This work was partly supported by
  Science Foundation Ireland under grants 13/ERC/12561
  and 12/RC/2278 (AMBER). Gavin D’Arcy was supported by INTEL (Irl.) Ltd. as a researcher in
  residence at CRANN.
  This project has received funding from the European Union's Horizon 2020
  research and innovation programme under grant agreement No DLV-737038
  (TRANSPIRE).

\bibliography{exchange_final}
\end{document}